\documentclass[aps,prb,showpacs,twocolumn,superscriptaddress,floatfix]{revtex4}

\usepackage{graphicx}% Include figure files
\usepackage{dcolumn}% Align table columns on decimal point
\usepackage{bm}% bold math
\usepackage{amsmath}
\usepackage{amssymb}

\begin{document}

\renewcommand{\vec}[1]{\mbox{\boldmath $#1$}}

%\preprint{}

\title{Emergence of Magnetic Long-range Order in Frustrated Pyrochlore Nd$_2$Ir$_2$O$_7$ with Metal-insulator Transition}


\author{K. Tomiyasu}
\email[Electronic address: ]{tomiyasu@m.tohoku.ac.jp}
\affiliation{Department of Physics, Tohoku University,
Aoba, Sendai 980-8578, Japan}
\author{K. Matsuhira}
\affiliation{Faculty of Engineering, Kyusyu Institute of Technology, 
Kitakyusyu 804-8550, Japan}
\author{K. Iwasa}
\affiliation{Department of Physics, Tohoku University,
Aoba, Sendai 980-8578, Japan}
\author{M. Watahiki}
\affiliation{Department of Physics, Tohoku University,
Aoba, Sendai 980-8578, Japan}
\author{S. Takagi}
\affiliation{Faculty of Engineering, Kyusyu Institute of Technology, 
Kitakyusyu 804-8550, Japan}
\author{M. Wakeshima}
\affiliation{Division of Chemistry, Graduate School of Science, 
Hokkiado University, Sapporo 060-0810, Japan}
\author{Y. Hinatsu}
\affiliation{Division of Chemistry, Graduate School of Science, 
Hokkiado University, Sapporo 060-0810, Japan}
\author{M. Yokoyama}
\affiliation{Faculty of Science, Ibaraki University,
Mito, Ibaraki 310-8512, Japan}
\author{K. Ohoyama}
\affiliation{Institute of Materials Research, Tohoku University,
Aoba, Sendai 980-8577, Japan}
\author{K. Yamada}
\affiliation{WPI Advanced Institute of Materials Research, Tohoku University,
Aoba, Sendai 980-8577, Japan}
%\thanks{}


\date{\today}

\begin{abstract}
In this study, we performed powder neutron diffraction and inelastic scattering measurements of frustrated pyrochlore Nd$_2$Ir$_2$O$_7$, which exhibits a metal-insulator transition at a temperature $T_{\rm MI}$ of 33 K.
The diffraction measurements revealed that the pyrochlore has an antiferromagnetic long-range structure with propagation vector $\vec{q}_{0}$ of (0,0,0) and that it grows with decreasing temperature below 15 K. This structure was analyzed to be of the all-in all-out type, consisting of highly anisotropic Nd$^{3+}$ magnetic moments of magnitude $2.3\pm0.4$$\mu_{\rm B}$, where $\mu_{\rm B}$ is the Bohr magneton. The inelastic scattering measurements revealed that the Kramers ground doublet of Nd$^{3+}$ splits below $T_{\rm MI}$. This suggests the appearance of a static internal magnetic field at the Nd sites, which probably originates from a magnetic order consisting of Ir$^{4+}$ magnetic moments.
Here, we discuss a magnetic structure model for the Ir order and the relation of the order to the metal-insulator transition in terms of frustration.
\end{abstract}

%\keywords{geometrical frustration, inelastic neutron scattering, spin molecule, dimer}

\maketitle

%
%%%
\section{Introduction}
%%%
%

%Geometrical frustration
Since Pauling's initial proposal for water ice,~\cite{Pauling_1935} geometrical frustration has played an important role in the fields of both chemistry and solid-state physics. Only some classical-spin pairs in frustrated magnets can be arranged antiferromagnetically on a triangular lattice.~\cite{Wannier_1950,Anderson_1956} Therefore, frustration suppresses magnetic ordering and promotes intriguing phenomena in insulators, such as spin-liquid-like fluctuations of spin ice and spin molecules,~\cite{Bramwell_2001,Lee_2002} and the formation of complex magnetic structures with multiferroics.~\cite{Tomiyasu_2004,Yamasaki_2006}

%Conductive frustrated systems
Metallic frustrated systems can exhibit novel transport properties. Examples of such properties include heavy fermion behaviors in spinel LiV$_2$O$_4$ and in the C15 Laves phase Y(Sc)Mn$_2$,~\cite{Matsushita_2005,Shiga_1993} superconductivity in pyrochlore oxide Cd$_2$Re$_2$O$_7$,~\cite{Hanawa_2001} anomalous Hall effects in pyrochlores Nd$_2$Mo$_2$O$_7$ and Pr$_2$Ir$_2$O$_7$,~\cite{Taguchi_2001,Machida_2007} metal-insulator (MI) transitions in pyrochlores Cd$_2$Os$_2$O$_7$ and Hg$_2$Ru$_2$O$_7$, and so on.~\cite{Sleight_1974,Yamamoto_2007} All these systems consist of corner-sharing tetrahedral lattices of magnetic atoms, called pyrochlore lattices, which provide an ideal platform for the occurrence of frustration. Thus, one of the challenging issues in current condensed matter physics is to determine the relation between the transport properties and geometrical frustration.

%Ln2Ir2O7: Lanthanide contraction and T_MI
Pyrochlore iridates $R_{2}$Ir$_2$O$_7$ ($R =$ Nd, Sm, Eu, Gd, Tb, Dy, or Ho) also exhibit MI transitions,~\cite{Yanagishima_2001,Matsuhira_2007,Matsuhira_2011} wherein both $R$ and Ir atoms form pyrochlore sublattices. All the MI transitions are of the second order.~\cite{Matsuhira_2007,Matsuhira_2011} The transition temperature $T_{\rm MI}$ monotonically increases from 33 to 141 K with increasing atomic number of the $R$ element from Nd to Ho.~\cite{Matsuhira_2007,Matsuhira_2011} The $R$ dependence of $T_{\rm MI}$ suggests that the MI transitions are sensitive to the change in the ionic radius of $R^{3+}$ (i.e. lanthanide contraction) but are independent of the 4$f$ electronic states.~\cite{Matsuhira_2007,Matsuhira_2011}

%Ln2Ir2O7: magnetism
Magnetic susceptibility measurements of $R_{2}$Ir$_2$O$_7$ in a previous study revealed that the zero-field-cooled and field-cooled curves do not follow each other below $T_{\rm MI}$.~\cite{Matsuhira_2007,Matsuhira_2011} This observation suggests that the MI transitions are accompanied by a magnetic anomaly.~\cite{Matsuhira_2007,Matsuhira_2011} Band calculations of the iridate in another study revealed that the conduction band consists of Ir 5$d$ and O 2$p$ orbitals in the metallic phase.~\cite{Koo_1998} Therefore, the magnetic anomaly seems to originate from the Ir magnetic moments. The Ir atoms are expected to be tetravalent magnetic ions in the insulating phase, as in the case of insulating Sr$_2$IrO$_4$ (Ir$^{4+}$: 5$d^5$, $J_{\rm eff}=1/2$, $L_{\rm eff}=1$, low-spin state $S=1/2$); here, $\vec{J}_{\rm eff}$ is the effective total angular momentum, $\vec{L}_{\rm eff}$ is the effective orbital momentum to describe triply degenerated $t_{2g}$ states under spin-orbit coupling, $\vec{S}$ is the spin angular momentum, and $\vec{J}_{\rm eff}=\vec{L}_{\rm eff}+\vec{S}$.~\cite{Kim_2008}

%Nd2Ir2O7
Recently, several microscopic experiments have further investigated Nd$_2$Ir$_2$O$_7$, and almost ensured that its MI transition originates from magnetic ordering at $T_{\rm MI}$. 
For example, Raman scattering experiments of Nd$_2$Ir$_2$O$_7$ suggested that there is no structural phase transition down to 4.2 K and that lattice distortion does not interestingly cause the MI transition.~\cite{Hasegawa_2010}
Further, in studies of temperature variation of the Nd 4$f$-electron states across $T_{\rm MI}$, crystalline electric field excitations of Nd$^{3+}$ were observed by inelastic neutron scattering, specific heat measurements, and magnetic susceptibility measurements, and the results were used to deduce the Ir magnetic states.~\cite{Watahiki_2011,Matsuhira_2011} The ten states of Nd$^{3+}$ (4$f^3$, $J=9/2$, $L=6$, $S=3/2$) split into five Kramers doublets. The first excited doublet exists at around 26 meV (300 K), which is much higher than $T_{\rm MI}$. The ground doublet possesses in/out-type Ising moments with 2.37$\mu_{\rm B}$ along the $\langle111\rangle$ direction, where $\mu_{\rm B}$ is the Bohr magneton. With decreasing temperature, the ground doublet energetically splits at $T_{\rm MI}$ like an order parameter of the MI transition, and the splitting energy is about 1.3 meV at 3 K. This phenomenon suggests the appearance of a static internal magnetic field at the Nd sites, probably caused by the magnetic ordering of Ir at $T_{\rm MI}$. 

%Summary of motivation
Thus, it is most probable that Nd$_2$Ir$_2$O$_7$ simultaneously undergoes the MI transition and the magnetic ordering without structural phase transition. However, the following major questions are still open: how the frustration is released without structural transition and how the MI transition is related to its release. To clarify them, information of the magnetic structure, which is prime for the study of magnetic frustration, will be necessary. 

%This paper
In this study, we conducted neutron diffraction and inelastic scattering experiments of Nd$_2$Ir$_2$O$_7$ in powder form. We also studied a magnetic structure model both for the Nd and Ir moments below $T_{\rm MI}$ and determined its relation to the MI transition in terms of frustration.

%
%%%
\section{Experimental}
%%%
%

%neutron instruments
Neutron diffraction experiments were performed on the powder diffractometer HERMES (T1-3) at the Institute for Materials Research (IMR), Tohoku University, installed in the thermal guide tube of the JRR-3 reactor at the Japan Atomic Energy Agency (JAEA).~\cite{HERMES_1998} Incident neutrons with an initial energy of 24.5 meV ($\lambda=1.8204(5)$ {\AA}) were extracted by a (331) reflection of a Ge monochromator and horizontal collimation sequence of guide-blank-sample-22$^\prime$.
Elastic neutron scattering experiments were performed on the triple-axis spectrometer TOPAN (6G) of Tohoku University, also installed in the JRR-3 reactor. The final energy of the neutrons was fixed to 30.5 meV with a horizontal collimation sequence of blank-60$^{\prime}$-sample-60$^{\prime}$-blank. A sapphire filter and a pyrolytic graphite filter efficiently removed fast neutrons and higher-order contamination, respectively.

Inelastic neutron scattering experiments were performed on the triple-axis spectrometer HER (C1-1) at the Institute for Solid State Physics (ISSP), University of Tokyo, installed in the cold guide tube of the JRR-3 reactor. The final energy of the neutrons was fixed to 3.6 meV with a horizontal collimation sequence of guide-blank-sample-radial-blank, where the radial collimator has three blank channels. A horizontal focusing analyzer, which covers the range of scattering angles by 5$^{\circ}$ and permits acquisition of higher statistics of data, was used. A cooled Be filter and a pyrolytic graphite Bragg-reflection filter efficiently removed the half-lambda contamination. The energy resolution (full width at half maximum) was 0.13 meV under the elastic condition.

%sample
A powder sample of Nd$_2$Ir$_2$O$_7$ was synthesized by a solid-state reaction method.~\cite{Matsuhira_2007,Matsuhira_2011} The sample (4.5 g) was placed in a thin aluminum foil and shaped to a hollow cylinder with a thickness of 0.7 mm and diameter of 20 mm in order to mitigate the effect of the strong neutron absorption of Ir nuclei as much as possible. Then, the cylinder was kept in an aluminum container that was placed under a cold head in a $^4$He or $^3$He closed-cycle refrigerator.

%
%%%
\section{Results}
%%%
%

\begin{figure}[htbp]
\begin{center}
\includegraphics[width=3.0in, keepaspectratio]{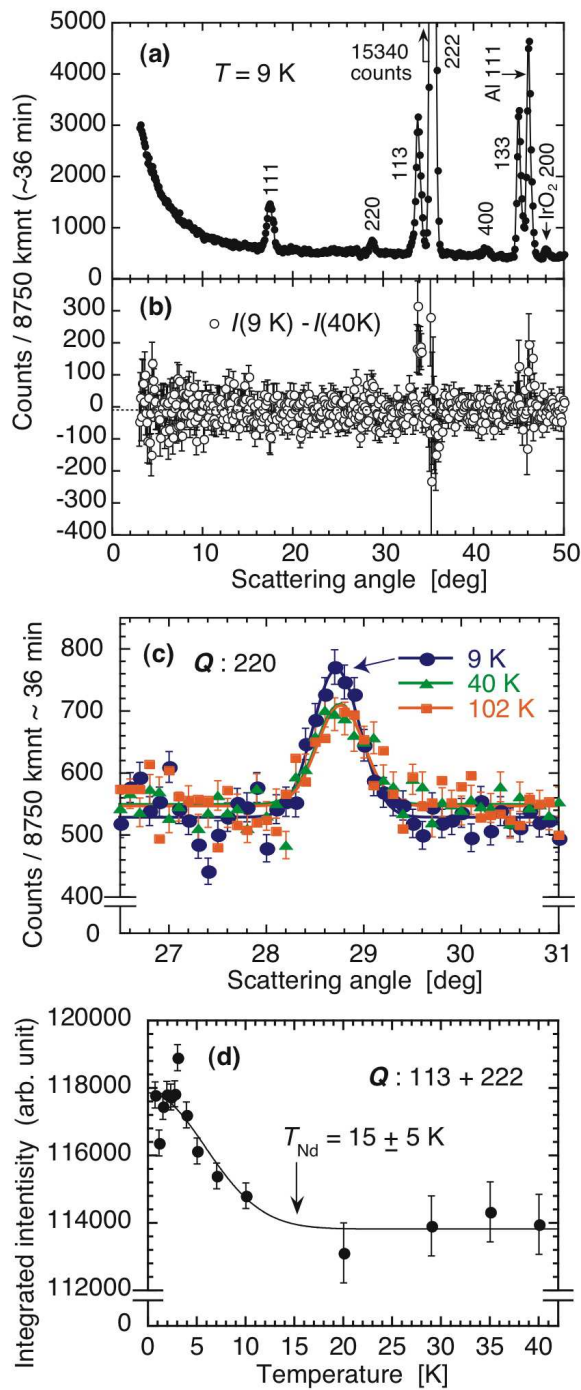}
\caption{\label{fig:HERMES} (Color online) Measured neutron diffraction data for powder Nd$_2$Ir$_2$O$_7$.
(a) Diffraction data measured at 9 K. (b) Diffraction data obtained by subtracting the 40 K data from the 9 K data (a). (c) Bragg reflection lines measured around the (220) reciprocal lattice point at 9 K, 40 K, and 102 K. (d) Temperature dependence of summation of the integrated intensities 113 and 222. In (a), (c), and (d), all the lines are a guide to the eye. }
\end{center}
\end{figure}

%Diffraction data
Figure~\ref{fig:HERMES}(a) shows the diffraction pattern measured at 9 K, i.e. below $T_{\rm MI}$.
In this figure, only fundamental reflections of the pyrochlore structure and a tiny peak of the IrO$_2$ impurity phase are observed; no superstructure lines are detected.
Figure~\ref{fig:HERMES}(b) shows the difference between diffraction data measured below and above $T_{\rm MI}$, i.e. at 9 K and 40 K, respectively. No appreciable magnetic diffuse scattering is observed, unlike that often seen as the spin-ice state in magnetic pyrochlore oxides.~\cite{Kadowaki_2002}
In contrast, a component characterized by the propagation vector $\vec{q}_{0}=(0,0,0)$ was observed below $T_{\rm MI}$. Figure~\ref{fig:HERMES}(c) shows the (220) fundamental reflection lines measured below and above $T_{\rm MI}$. A weak yet noticeable increase in the intensity is observed as the temperature decreases from above $T_{\rm MI}$ to below it, and no broadening of the line width is observed. These results suggest a long-range structure of the pyrochlore with $\vec{q}_{0}$. Further, since previous Raman scattering experiments have shown that structural transition is not observed down to 4.2 K,~\cite{Hasegawa_2010} the $\vec{q}_{0}$ structure is most probably magnetic. 

Table~\ref{tab:mag} summarizes the difference between integrated intensities at 40 K and 9 K for several fundamental reflections in barn/unit cell, where a unit cell refers to a crystallographic unit cell denoted by 2(Nd$_2$Ir$_2$O$_{7}$). The scale conversion factor for the units was estimated by analyzing the fundamental nuclear reflection intensities at 40 K, where the lattice parameter $x=0.33(0)$ coinciding with those for isomorphic materials Sm$_2$Ir$_2$O$_7$ and Eu$_2$Ir$_2$O$_7$ as estimated by room-temperature X-ray diffraction was obtained.~\cite{Taira_2001, comm_1} 
The conversion factor also includes the absorption correction, since the correction is canceled in the ratio to the nuclear reflection intensities. Further, the reflection-angle dependence of the correction can be approximately ignored, since the sample shape is cylindrical. We confirmed that the nuclear reflection intensities at 40 K are well analyzed without the reflection-angle dependence of the correction. 

%In addition, as will be discussed in {\S} 4, the increments can be explained using a magnetic structure model.
%Also, the intensity increments overall decrease with increasing $hkl$ indices, and no appreciable difference was obtained at $hkl$ indices higher than 135 (scattering angle $2\theta > 63$ deg).
%
\begin{table}[htbp]
\begin{center}
\caption{\label{tab:mag} Integrated intensities of magnetic components of several fundamental reflections in barn/unit cell. Here, a unit cell refers to a crystallographic unit cell denoted by 2(Nd$_2$Ir$_2$O$_{7}$). The experimental intensities were evaluated by subtracting the integrated intensities at 40 K from those at 9 K. The best-fit calculated intensities were obtained for the all-in all-out structure shown in Figs.~\ref{fig:model}(b) and \ref{fig:model}(c). The experimental and calculated intensities are compared in Fig.~\ref{fig:model}(a) as well.
%The $R$ factor is 38 {\%}.
}
\bigskip
%\begin{ruledtabular}
\begin{tabular}{cccc}
\hline
$hkl$ & Exp. & Cal. (Nd only) & Cal. (Nd and Ir) \\
\hline
111 & $-5.9\pm5.4$ & 0 & 0\\
200 & $0\pm2.5$ & 0 & 0\\
220 & $13.6\pm3.5$ & 16.8 & 13.7\\
113 + 222 & $23.3\pm13.7$ & 17.4 &20.6\\
400 & $-2.4\pm3.9$ & 0 & 0\\
331 & $5.0\pm5.8$ & 5.5 & 6.4\\
420 & $9.3\pm6.7$ & 9.8 & 8.1\\
422 & $0.9\pm4.5$ & 3.3 & 2.7\\
135 & $4.9\pm2.9$ & 8.6 & 10.0\\
\hline
\end{tabular}
%\end{ruledtabular}
\end{center}
\end{table}
%

\begin{figure}[htbp]
\begin{center}
\includegraphics[width=3.0in, keepaspectratio]{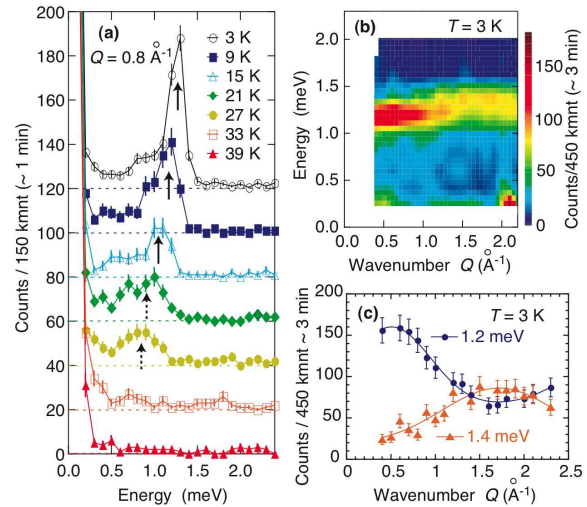}
\caption{\label{fig:HER} (Color online) Measured inelastic neutron scattering raw data for powder Nd$_2$Ir$_2$O$_7$. (a) Scan data measured at constant $Q$ of 0.8 {\AA}$^{-1}$ at different temperatures.~\cite{Watahiki_2011} The vertical origins shift to the height indicated by the horizontal dotted lines. (b) Inelastic scattering intensity distributions in $(Q,E)$ space at 3 K. (c) Comparison results of Constant-$E$ scans measured at $E=1.2$ and 1.4 meV in (b). In (a) and (c), the solid curves are a guide to the eye. }
\end{center}
\end{figure}

%Diffraction & inelastic data
Figure~\ref{fig:HERMES}(d) shows the temperature dependence of the summation of the integrated intensities 113 and 222; this summation shows the largest intensity increment (see Table~\ref{tab:mag}). With decreasing temperature, the intensity begins to increase at around $15\pm5$ K, and not $T_{\rm MI}$.
A similar temperature dependence is observed in the inelastic scattering data. Figure~\ref{fig:HER}(a) shows the energy ($E$) spectra at a fixed scattering wavenumber ($Q$).
The following behavior is observed with decreasing temperature. A broad inelastic signal appears at around $T_{\rm MI}$, as shown by the dotted arrows; it changes into a sharp peak at 1.1 meV at around 15 K; below 15 K, the peak sharpens further, as shown by the solid arrows; in the previous study, it was not important for us to know at what temperature the peak began sharpening.~\cite{Watahiki_2011} The emergence of the sharp inelastic peak can be attributed to the splitting of the ground doublet of Nd 4$f$ electrons.~\cite{Watahiki_2011} The agreement between the data in Figs.~\ref{fig:HERMES}(d) and \ref{fig:HER}(a) strongly suggests that the $\vec{q}_{0}$ structure mainly originates from Nd$^{3+}$ magnetic moments.
%consistent with thermal population

%Inelastic data
Figure~\ref{fig:HER}(b) shows the inelastic scattering intensity distributions in $(Q,E)$ space measured below $T_{\rm MI}$. The horizontally spreading excitation mode is observed at around 1.3 meV; it corresponds to the splitting of the Nd$^{3+}$ ground doublet. The scattering intensity decreases with increasing $Q$, confirming that the excitations are magnetic. However, the mode appears slightly dispersive. To verify the dispersion, we compared the constant-$E$ scans measured at 1.2 and 1.4 meV in Fig.~\ref{fig:HER}(b); the comparison results are shown in Fig.~\ref{fig:HER}(c). The $Q$ dependences at these energies are clearly different, indicating that the excitations are not completely flat but dispersive, with a bandwidth of $\sim$ 0.1 meV (1 K).

%
%%%
\section{Analyses}
%%%
%

%Magnetic structure analysis
%Only Nd$^{3+}$
We analyzed the magnetic structure on the basis of the $\vec{q}_{0}$ intensities given in Table~\ref{tab:mag}. Crystalline field analysis~\cite{Watahiki_2011} revealed the magnitude of Nd$^{3+}$ moments to be about 2.37$\mu_{\rm B}$, whereas that of Ir$^{4+}$ moments is expected to be 1$\mu_{\rm B}$ at most. Therefore, as the first approximation, we assumed that the $\vec{q}_{0}$ intensities consist of only the Nd moments. In fact, the statistical errors of our data would be too large to resolve the Nd and Ir moments. Furthermore, later in this section, the magnitude of moment estimated by the magnetic structure analysis is confirmed to be in agreement with that estimated by the crystalline field analysis.

\begin{figure*}[htbp]
\begin{center}
\includegraphics[width=0.95\linewidth, keepaspectratio]{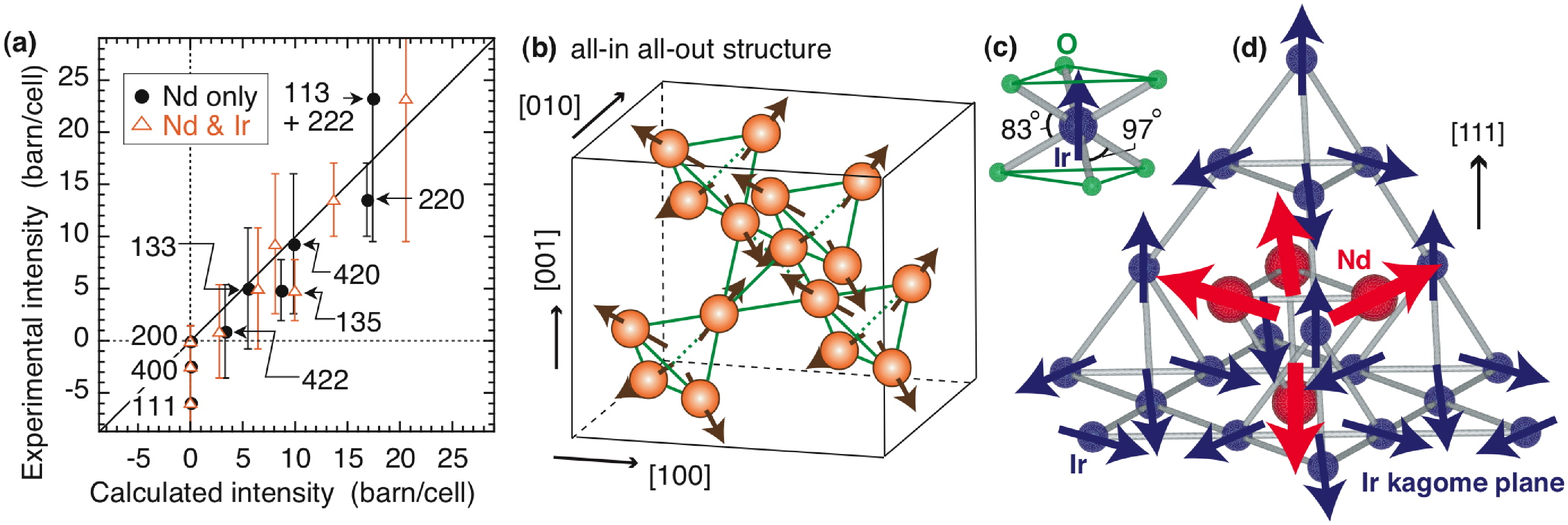}
\caption{\label{fig:model} (Color online) Magnetic structure modelling.
(a) Comparison between experimental and best-fit calculated magnetic reflection intensities given in Table~\ref{tab:mag}. 
(b) All-in all-out magnetic structure. The filled circles represent Nd$^{3+}$ ions, and the arrows represent the Nd moments.
(c) Trigonally distorted O$^{2-}$ ligand around Ir$^{4+}$ ion. All the Ir-O bonds are of the same length (2.01 {\AA}).
(d) Relation between magnetic moments (blue thin arrows) of Ir$^{4+}$ ions (blue small balls) and magnetic moments (red thick arrows) of Nd$^{3+}$ ions (red large balls). Both the Ir and the Nd moments form the all-in all-out structures. Alternative directions of Nd moments in the case of a ferromagnetic Nd-Ir interaction are depicted: when the moments are antiferromagnetic, the directions of all the red arrows are reversed, and the all-in all-out type structure is retained. }
\end{center}
\end{figure*}

%Only in/out type
The crystalline field analysis also strongly suggests that the Nd moments are highly anisotropic along the $\langle111\rangle$ directions (in/out-type Ising moments),~\cite{Watahiki_2011} as expected from available data on other Nd pyrochlore oxides.~\cite{Yasui_2001} In addition, the measured magnetic susceptibility shows a small value of only $\sim 10^{-3}$$\mu_{\rm B}$/formula under a magnetic field of 1 kOe, and no hysteresis curve is observed at 5 K.~\cite{Taira_2001,Matsuhira_2007,Watahiki_2011,Matsuhira_2011} Therefore, among the various magnetic structure models described by the in/out-type moments, the 3-in 1-out and the 2-in 2-out types are most probably ruled out, because these model types are essentially ferromagnetic. A unique possible solution is the all-in all-out type of model, as shown in Fig.~\ref{fig:model}(b). Thus, we examined the consistency between the all-in all-out model and the $\vec{q}_{0}$ intensities.

%Fitting of the data at 9 K
The remaining fitting parameter is the absolute value of magnetic moments of Nd$^{3+}$ ($m$). Therefore, we evaluated it by the least-squares method, by employing the magnetic form factor of Nd$^{3+}$ calculated by Freeman and Desclaux.~\cite{Freeman_1979} Since the sample was cylindrical in shape, the reflection-angle dependence of the absorption factor was ignored. The evaluation results showed that the all-in all-out model had the best-fit calculated intensities, with $m(9 \phantom{0} {\rm K}) = 1.3\pm0.2$ $\mu_{\rm B}$ (Table~\ref{tab:mag}); these intensities are in agreement with the experimental ones, as shown in Fig.~\ref{fig:model}(a).

%Moments at 0.7 K
Further, from Fig.~\ref{fig:HERMES}(d), the ratio of magnetic scattering intensity at 0.7 K to that at 9 K is estimated to be about 3.2. Since the magnetic intensity is proportional to the square of $m$,~\cite{Marshall_1971} the value of $m(0.7 \phantom{0} {\rm K})$ is estimated to be $2.3\pm0.4$ (=$\sqrt{3.2} \times m(9 \phantom{0} {\rm K})$)$\mu_{\rm B}$. This value is in good agreement with the value of 2.37$\mu_{\rm B}$ estimated by the crystalline field analysis.~\cite{Watahiki_2011}
%

%Summary of RESULTS and ANALYSIS
To summarize Secs. 3 and 4, we found that the antiferromagnetic long-range structure with $\vec{q}_0$ grows with decreasing temperature below $T_{\rm Nd} = 15 \pm 5$ K. The structure can be approximately described by the all-in all-out type of model for Nd moments with a magnitude of $2.3\pm0.4$$\mu_{\rm B}$ at 0.7 K. No direct signals for Ir moments were obtained in the present experiments.

%
%%%
\section{Discussion}
%%%
%

%
\subsection{Magnetic structure below $T_{\rm MI}$}
%

%Existence of hidden Ir order
The ground doublet of Nd$^{3+}$ splits when the temperature decreases to below $T_{\rm MI}$, as shown by the dotted arrows in Fig.~\ref{fig:HER}(a); however, the Nd magnetic structure grows mainly below $T_{\rm Nd}$, as shown in Fig.~\ref{fig:HERMES}(d) and by the solid arrows in Fig.~\ref{fig:HER}(a). This discrepancy in temperature suggests that not Nd moments exhibit static magnetic order at $T_{\rm MI}$. In isomorphic Nd$_2$Mo$_2$O$_7$, the magnetic moments of $d$ and $f$ electrons do not order simultaneously, as revealed by neutron diffraction experiments; the Mo structure grows below Curie temperature $T_{C}=93$ K and then the Nd structure grows mainly below $\sim20$ K with decreasing temperature.~\cite{Yasui_2001} In analogy with Nd$_2$Mo$_2$O$_7$, Ir magnetic ordering is expected to occur in Nd$_2$Ir$_2$O$_7$ at $T_{\rm MI}$. The ordered Ir moment is estimated to be as small as the errors of $m$ later in this subsection.

The splitting of the ground doublet of Nd$^{3+}$ is slightly dispersive, with a width of 0.1 meV (1 K), as shown in Fig.~\ref{fig:HER}(b); this width roughly corresponds to the magnitude of Nd-Nd interactions. The isomorphic pyrochlore insulator Nd$_2$Sn$_2$O$_7$ with a similar magnitude of Nd$^{3+}$ magnetic moments (2.6$\mu_{\rm B}$) also exhibits a Ne$\grave{\rm e}$l temperature of 0.9 K and a Curie-Weiss temperature of $-0.3$ K,~\cite{Matsuhira_2002} again indicating that the magnitude is about 1 K. Thus, the Nd structure growing below $T_{\rm Nd}$ is not formed by Nd-Nd interactions alone. The structure probably parasitizes the aforementioned hidden Ir order through Nd-Ir exchange interactions, whose magnitude is estimated to be about the splitting width of 1.3 meV $\sim$ 15 K; this width is in excellent agreement with the value of $T_{\rm Nd}$.

%Correlation of Ir order (spin-ice type frustration)
We next discuss the Ir magnetic structure. First, as shown in Fig.~\ref{fig:HER}(a), a single sharp peak appears at 3 K. This fact indicates that the internal magnetic field is uniform at all Nd$^{3+}$ sites in the entire sample; this uniformity in turn suggests that the Ir structure is also described by $\vec{q}_{0}$ without any modulation. Indeed, $\vec{q}_{0}$ has no special symmetrical directions and therefore forms no domains, and is consistent with the absence of structural transition. Second, as mentioned above, the antiferromagnetic nature of the magnetic structure is clear from the magnetic susceptibility. Third, as shown in Fig.~\ref{fig:model}(c), the local environment of Ir$^{4+}$ is trigonal, suggesting that Ir moments with $J_{\rm eff}=1/2$ also favour the $\langle111\rangle$ directions (in/out type). Thus, from these three observations, along the same lines as for the Nd structure mentioned in {\S} 4, the Ir moments are also expected to form the all-in all-out type of antiferromagnetic structure.

We can confirm that the all-in all-out type of Ir structure generates the all-in all-out type of Nd structure, when only nearest-neighbor Nd-Ir exchange interactions are taken into account. Figure~\ref{fig:model}(d) shows the relation of Ir and Nd crystallographic sites and magnetic moments with the all-in all-out structures. The single Nd site is positioned at the centre of the hexagon on the kagome plane in the Ir sublattice. At the Nd site, the surrounding six nearest-neighbor Ir moments generate the exchange field, which is perpendicular to the (111) kagome plane. The alternative upward or downward direction depends on whether the Nd-Ir interaction is ferromagnetic or antiferromagnetic.

To examine the consistency between the approximation with only Nd moments in {\S} 4 and the aforementioned model with both Nd and Ir structures, it will be worth reanalyzing the experimental reflection intensities (Table~\ref{tab:mag}), though the statistical errors are large. The fitting parameters are the absolute value of magnetic moments of Nd$^{3+}$ ($m_{\rm Nd}>0$) and that of Ir$^{4+}$ ($m_{\rm Ir}$) with sign corresponding to the alternative upward or downward direction. We evaluated them by the least-squares method, by employing the magnetic form factors of Nd$^{3+}$ and Ir$^{4+}$.~\cite{Freeman_1979, Kobayashi_2011} The best-fit calculated intensities (Table~\ref{tab:mag}) were obtained at $m_{\rm Nd}(9 \phantom{0} {\rm K}) = 1.3$ $\mu_{\rm B}$ and $m_{\rm Ir}(9 \phantom{0} {\rm K}) = -0.1$ $\mu_{\rm B}$, which correspond to $m_{\rm Nd}(0.7 \phantom{0} {\rm K}) = 2.3$ $\mu_{\rm B}$ and $m_{\rm Ir}(0.7 \phantom{0} {\rm K}) = -0.2$ $\mu_{\rm B}$. The negative sign means the magnetic structure shown in Fig.~\ref{fig:model}(c) (ferromagnetic Nd-Ir interaction). These calculated intensities are again in agreement with the experimental ones within the statistical errors, as shown in Fig.~\ref{fig:model}(a). Thus, the very small value of $m_{\rm Ir}$ supports the validity of the approximation with only Nd moments.

%The magnitude of ordered Ir moment is unknown. Since low-energy magnetic excitations weakly survive in the energy spectrum at 3 K in the left side of the sharp peak, as shown in Fig.~\ref{fig:HER}(a), magnetic fluctuations possibly coexist with the Ir and Nd magnetic structures. We also remark that muon spin resonance experiments suggested the existence of quasi-static magnetic components in another pyrochlore oxide with an MI transition, Cd$_2$Os$_2$O$_7$.~\cite{Koda_2007}

%
\subsection{Relation between frustration and MI transition}
%

%frustration
The present experiments revealed that the antiferromagnetic long-range structure with $\vec{q}_0$ exists below $T_{\rm MI}$ in Nd$_2$Ir$_2$O$_7$. This fact strongly suggests that the frustration is released below $T_{\rm MI}$. That is, the system is expected to change from {\it the magnetically frustrated metallic phase} to {\it the antiferromagnetic insulating phase} at $T_{\rm MI}$. The metallic phase will gain orbital hybridization energy (band formation energy), whereas the insulating phase will gain magnetic ordering energy.

Then, the following question arises: how is the frustration released without lattice distortion at $T_{\rm MI}$? In a pyrochlore lattice, antiferromagnetic nearest-neighbor interactions with {\it isotropic} moments induce strong frustration, as is the case in the highly frustrated spinel MgCr$_2$O$_4$ (Cr$^{3+}$: 3$d^3$),~\cite{Tomiyasu_2008} whereas those with {\it anisotropic} moments induce the all-in all-out type long-range order without frustration.~\cite{Maegawa_2010,comm_2} The anisotropy is based on the spin-orbit-coupled $\vec{J}_{\rm eff}$ states. The spin-orbit coupling is normally suppressed in a metallic phase (unlike in an insulating phase), since the coupling constant $\lambda$ is roughly proportional to $1/r^{3}$, where $r$ is the size of an unpaired electron cloud.
Thus, in Nd$_2$Ir$_2$O$_7$, above $T_{\rm MI}$, an antiferromagnetic interaction with relatively isotropic Ir moments is considered to hamper magnetic ordering owing to frustration. Below $T_{\rm MI}$, the spin-orbit coupling will become relatively active and the Ir moments will restore the in/out-type anisotropy; therefore, the all-in all-out type of order will emerge without frustration.

%Band gap
It is open how the electron/hole bands change below and above $T_{\rm MI}$. We remark that the Mott-type antiferromagnetic insulator with $J_{\rm eff}=1/2$ half-filled bands was recently established in Sr$_2$IrO$_4$.~\cite{Kim_2008,Kim_2009} The $J_{\rm eff}=1/2$ Mott insulator might be related to the insulating phase in Nd$_2$Ir$_2$O$_7$.

%
%%%
\section{Conclusions}
%%%
%

%Nd magnetic structure
In summary, to study a relation among magnetic frustration, the absence of structural phase transition, and the MI transition in Nd$_2$Ir$_2$O$_7$, we conducted neutron experiments of this material in powder form. As the results, magnetic Bragg reflections with $\vec{q}_{0}=(0,0,0)$ were found by diffraction. The propagation vector is consistent with the absence of structural transition. The reflection intensity of Nd$_2$Ir$_2$O$_7$ increases with decreasing temperature below $T_{\rm Nd} = 15 \pm 5$ K. The magnetic structure of this system can be described by the all-in all-out type of model for Nd moments with a magnitude of about $2.3\pm0.4$$\mu_{\rm B}$ at 0.7 K. The magnitude of moments is consistent with the results of a previous crystalline field analysis.

%hidden Ir magnetic order
The Kramers ground doublet of Nd$^{3+}$ begins to split at $T_{\rm MI}$ with decreasing temperature, suggesting another magnetic ordering at this temperature. The magnitude of dispersion of the splitting is estimated to be only $\sim$ 1 K, which is much below $T_{\rm Nd}$; this suggests that the Nd magnetic structure cannot be formed by Nd-Nd interactions alone. Therefore, we concluded that a hidden magnetic order of Ir moments appears at $T_{\rm MI}$, and it is parasitized by the Nd structure; however, we could not detect direct signals for the Ir moments in the experiments, probably because of their very small magnitude.

%Ir structure and its relation to MI transition
The Ir magnetic structure can also be described by the all-in all-out type of structure. We proposed a scenario that the in/out type of anisotropy based on spin-orbit coupling plays a key role in releasing the frustration and inducing the MI transition. The scenario will provide an example of the relation between the transport properties and geometrical frustration. 

\acknowledgments
We thank Mr. M. Ohkawara, Mr. K. Nemoto, and Mr. T. Asami for providing assistance at the JAEA, and Mr. M. Onodera for providing assistance at Tohoku University. The neutron experiments at the JAEA were performed under User Programs conducted by ISSP. This study was financially supported by Grants-in-Aid for Young Scientists (B) (22740209); Priority Areas (22014001 and 19052005); Scientific Researches (C) (23540417), (S) (21224008), and (A) (22244039); and Innovative Areas (20102005 and 21102518) from the MEXT of Japan. The study was also supported by the Inter-university Cooperative Research Program of the Institute for Materials Research at Tohoku University. 

\bibliography{65141}

\begin{thebibliography}{35}
\expandafter\ifx\csname natexlab\endcsname\relax\def\natexlab#1{#1}\fi
\expandafter\ifx\csname bibnamefont\endcsname\relax
  \def\bibnamefont#1{#1}\fi
\expandafter\ifx\csname bibfnamefont\endcsname\relax
  \def\bibfnamefont#1{#1}\fi
\expandafter\ifx\csname citenamefont\endcsname\relax
  \def\citenamefont#1{#1}\fi
\expandafter\ifx\csname url\endcsname\relax
  \def\url#1{\texttt{#1}}\fi
\expandafter\ifx\csname urlprefix\endcsname\relax\def\urlprefix{URL }\fi
\providecommand{\bibinfo}[2]{#2}
\providecommand{\eprint}[2][]{\url{#2}}

\bibitem[{\citenamefont{Pauling}(1935)}]{Pauling_1935}
\bibinfo{author}{\bibfnamefont{L.}~\bibnamefont{Pauling}}, \bibinfo{journal}{J.
  Am. Chem. Soc.} \textbf{\bibinfo{volume}{57}}, \bibinfo{pages}{2680}
  (\bibinfo{year}{1935}).

\bibitem[{\citenamefont{Wannier}(1950)}]{Wannier_1950}
\bibinfo{author}{\bibfnamefont{G.~H.} \bibnamefont{Wannier}},
  \bibinfo{journal}{Phys. Rev.} \textbf{\bibinfo{volume}{79}},
  \bibinfo{pages}{357} (\bibinfo{year}{1950}).

\bibitem[{\citenamefont{Anderson}(1956)}]{Anderson_1956}
\bibinfo{author}{\bibfnamefont{P.~W.} \bibnamefont{Anderson}},
  \bibinfo{journal}{Phys. Rev.} \textbf{\bibinfo{volume}{102}},
  \bibinfo{pages}{1008} (\bibinfo{year}{1956}).

\bibitem[{\citenamefont{Bramwell and Gingras}(2001)}]{Bramwell_2001}
\bibinfo{author}{\bibfnamefont{S.~T.} \bibnamefont{Bramwell}} \bibnamefont{and}
  \bibinfo{author}{\bibfnamefont{M.~J.~P.} \bibnamefont{Gingras}},
  \bibinfo{journal}{Science} \textbf{\bibinfo{volume}{294}},
  \bibinfo{pages}{1495} (\bibinfo{year}{2001}).

\bibitem[{\citenamefont{Lee et~al.}(2002)\citenamefont{Lee, Broholm, Ratcliff,
  Gasparovic, Huang, Kim, and Cheong}}]{Lee_2002}
\bibinfo{author}{\bibfnamefont{S.-H.} \bibnamefont{Lee}},
  \bibinfo{author}{\bibfnamefont{C.}~\bibnamefont{Broholm}},
  \bibinfo{author}{\bibfnamefont{W.}~\bibnamefont{Ratcliff}},
  \bibinfo{author}{\bibfnamefont{G.}~\bibnamefont{Gasparovic}},
  \bibinfo{author}{\bibfnamefont{Q.}~\bibnamefont{Huang}},
  \bibinfo{author}{\bibfnamefont{T.~H.} \bibnamefont{Kim}}, \bibnamefont{and}
  \bibinfo{author}{\bibfnamefont{S.-W.} \bibnamefont{Cheong}},
  \bibinfo{journal}{Nature (London)} \textbf{\bibinfo{volume}{418}},
  \bibinfo{pages}{856} (\bibinfo{year}{2002}).

\bibitem[{\citenamefont{Tomiyasu et~al.}(2004)\citenamefont{Tomiyasu, Fukunaga,
  and Suzuki}}]{Tomiyasu_2004}
\bibinfo{author}{\bibfnamefont{K.}~\bibnamefont{Tomiyasu}},
  \bibinfo{author}{\bibfnamefont{J.}~\bibnamefont{Fukunaga}}, \bibnamefont{and}
  \bibinfo{author}{\bibfnamefont{H.}~\bibnamefont{Suzuki}},
  \bibinfo{journal}{Phys. Rev. B} \textbf{\bibinfo{volume}{70}},
  \bibinfo{pages}{214434} (\bibinfo{year}{2004}).

\bibitem[{\citenamefont{Yamasaki et~al.}(2006)\citenamefont{Yamasaki, Miyasaka,
  Kaneko, He, Arima, and Tokura}}]{Yamasaki_2006}
\bibinfo{author}{\bibfnamefont{Y.}~\bibnamefont{Yamasaki}},
  \bibinfo{author}{\bibfnamefont{S.}~\bibnamefont{Miyasaka}},
  \bibinfo{author}{\bibfnamefont{Y.}~\bibnamefont{Kaneko}},
  \bibinfo{author}{\bibfnamefont{J.-P.} \bibnamefont{He}},
  \bibinfo{author}{\bibfnamefont{T.}~\bibnamefont{Arima}}, \bibnamefont{and}
  \bibinfo{author}{\bibfnamefont{Y.}~\bibnamefont{Tokura}},
  \bibinfo{journal}{Phys. Rev. Lett.} \textbf{\bibinfo{volume}{96}},
  \bibinfo{pages}{207204} (\bibinfo{year}{2006}).

\bibitem[{\citenamefont{Matsushita et~al.}(2005)\citenamefont{Matsushita, Ueda,
  and Ueda}}]{Matsushita_2005}
\bibinfo{author}{\bibfnamefont{Y.}~\bibnamefont{Matsushita}},
  \bibinfo{author}{\bibfnamefont{H.}~\bibnamefont{Ueda}}, \bibnamefont{and}
  \bibinfo{author}{\bibfnamefont{Y.}~\bibnamefont{Ueda}},
  \bibinfo{journal}{Nature Mater.} \textbf{\bibinfo{volume}{4}},
  \bibinfo{pages}{843} (\bibinfo{year}{2005}).

\bibitem[{\citenamefont{Shiga et~al.}(1993)\citenamefont{Shiga, Fujisawa, and
  Wada}}]{Shiga_1993}
\bibinfo{author}{\bibfnamefont{M.}~\bibnamefont{Shiga}},
  \bibinfo{author}{\bibfnamefont{K.}~\bibnamefont{Fujisawa}}, \bibnamefont{and}
  \bibinfo{author}{\bibfnamefont{H.}~\bibnamefont{Wada}}, \bibinfo{journal}{J.
  Phys. Soc. Jpn} \textbf{\bibinfo{volume}{A62}}, \bibinfo{pages}{1329}
  (\bibinfo{year}{1993}).

\bibitem[{\citenamefont{Hanawa et~al.}(2001)\citenamefont{Hanawa, Muraoka,
  Tayama, Sakakibara, Yamaura, and Hiroi}}]{Hanawa_2001}
\bibinfo{author}{\bibfnamefont{M.}~\bibnamefont{Hanawa}},
  \bibinfo{author}{\bibfnamefont{Y.}~\bibnamefont{Muraoka}},
  \bibinfo{author}{\bibfnamefont{T.}~\bibnamefont{Tayama}},
  \bibinfo{author}{\bibfnamefont{T.}~\bibnamefont{Sakakibara}},
  \bibinfo{author}{\bibfnamefont{J.}~\bibnamefont{Yamaura}}, \bibnamefont{and}
  \bibinfo{author}{\bibfnamefont{Z.}~\bibnamefont{Hiroi}},
  \bibinfo{journal}{Phys. Rev. Lett.} \textbf{\bibinfo{volume}{87}},
  \bibinfo{pages}{187001} (\bibinfo{year}{2001}).

\bibitem[{\citenamefont{Taguchi et~al.}(2001)\citenamefont{Taguchi, Oohara,
  Yoshizawa, Nagaosa, and Tokura}}]{Taguchi_2001}
\bibinfo{author}{\bibfnamefont{Y.}~\bibnamefont{Taguchi}},
  \bibinfo{author}{\bibfnamefont{Y.}~\bibnamefont{Oohara}},
  \bibinfo{author}{\bibfnamefont{H.}~\bibnamefont{Yoshizawa}},
  \bibinfo{author}{\bibfnamefont{N.}~\bibnamefont{Nagaosa}}, \bibnamefont{and}
  \bibinfo{author}{\bibfnamefont{Y.}~\bibnamefont{Tokura}},
  \bibinfo{journal}{Science} \textbf{\bibinfo{volume}{2573}},
  \bibinfo{pages}{291} (\bibinfo{year}{2001}).

\bibitem[{\citenamefont{Machida et~al.}(2007)\citenamefont{Machida, Nakatsuji,
  Maeno, Tayama, Sakakibara, and Onoda}}]{Machida_2007}
\bibinfo{author}{\bibfnamefont{Y.}~\bibnamefont{Machida}},
  \bibinfo{author}{\bibfnamefont{S.}~\bibnamefont{Nakatsuji}},
  \bibinfo{author}{\bibfnamefont{Y.}~\bibnamefont{Maeno}},
  \bibinfo{author}{\bibfnamefont{T.}~\bibnamefont{Tayama}},
  \bibinfo{author}{\bibfnamefont{T.}~\bibnamefont{Sakakibara}},
  \bibnamefont{and} \bibinfo{author}{\bibfnamefont{S.}~\bibnamefont{Onoda}},
  \bibinfo{journal}{Phys. Rev. Lett.} \textbf{\bibinfo{volume}{98}},
  \bibinfo{pages}{057203} (\bibinfo{year}{2007}).

\bibitem[{\citenamefont{Sleight et~al.}(1974)\citenamefont{Sleight, Gilison,
  Weiher, and Bindloss}}]{Sleight_1974}
\bibinfo{author}{\bibfnamefont{A.~W.} \bibnamefont{Sleight}},
  \bibinfo{author}{\bibfnamefont{J.~L.} \bibnamefont{Gilison}},
  \bibinfo{author}{\bibfnamefont{J.~F.} \bibnamefont{Weiher}},
  \bibnamefont{and} \bibinfo{author}{\bibfnamefont{W.}~\bibnamefont{Bindloss}},
  \bibinfo{journal}{Solid State Commun.} \textbf{\bibinfo{volume}{14}},
  \bibinfo{pages}{357} (\bibinfo{year}{1974}).

\bibitem[{\citenamefont{Yamamoto et~al.}(2007)\citenamefont{Yamamoto, Sharma,
  Okamoto, Nakao, Aruga-Katori, Niitaka, Hashizume, and
  Takagi}}]{Yamamoto_2007}
\bibinfo{author}{\bibfnamefont{A.}~\bibnamefont{Yamamoto}},
  \bibinfo{author}{\bibfnamefont{P.~A.} \bibnamefont{Sharma}},
  \bibinfo{author}{\bibfnamefont{Y.}~\bibnamefont{Okamoto}},
  \bibinfo{author}{\bibfnamefont{A.}~\bibnamefont{Nakao}},
  \bibinfo{author}{\bibfnamefont{H.}~\bibnamefont{Aruga-Katori}},
  \bibinfo{author}{\bibfnamefont{S.}~\bibnamefont{Niitaka}},
  \bibinfo{author}{\bibfnamefont{D.}~\bibnamefont{Hashizume}},
  \bibnamefont{and} \bibinfo{author}{\bibfnamefont{H.}~\bibnamefont{Takagi}},
  \bibinfo{journal}{J. Phys. Soc. Jpn.} \textbf{\bibinfo{volume}{76}},
  \bibinfo{pages}{043703} (\bibinfo{year}{2007}).

\bibitem[{\citenamefont{Yanagishima and Maeno}(2001)}]{Yanagishima_2001}
\bibinfo{author}{\bibfnamefont{D.}~\bibnamefont{Yanagishima}} \bibnamefont{and}
  \bibinfo{author}{\bibfnamefont{Y.}~\bibnamefont{Maeno}}, \bibinfo{journal}{J.
  Phys. Soc. Jpn.} \textbf{\bibinfo{volume}{70}}, \bibinfo{pages}{2880}
  (\bibinfo{year}{2001}).

\bibitem[{\citenamefont{Matsuhira et~al.}(2007)\citenamefont{Matsuhira,
  Wakishima, Nakanishi, Yamada, Nakamura, Kawano, Takagi, and
  Hinatsu}}]{Matsuhira_2007}
\bibinfo{author}{\bibfnamefont{K.}~\bibnamefont{Matsuhira}},
  \bibinfo{author}{\bibfnamefont{M.}~\bibnamefont{Wakishima}},
  \bibinfo{author}{\bibfnamefont{R.}~\bibnamefont{Nakanishi}},
  \bibinfo{author}{\bibfnamefont{T.}~\bibnamefont{Yamada}},
  \bibinfo{author}{\bibfnamefont{A.}~\bibnamefont{Nakamura}},
  \bibinfo{author}{\bibfnamefont{W.}~\bibnamefont{Kawano}},
  \bibinfo{author}{\bibfnamefont{S.}~\bibnamefont{Takagi}}, \bibnamefont{and}
  \bibinfo{author}{\bibfnamefont{Y.}~\bibnamefont{Hinatsu}},
  \bibinfo{journal}{J. Phys. Soc. Jpn.} \textbf{\bibinfo{volume}{76}},
  \bibinfo{pages}{043706} (\bibinfo{year}{2007}).

\bibitem[{\citenamefont{Matsuhira et~al.}(to be
  published)\citenamefont{Matsuhira, Wakeshima, Hinatsu, and
  Takagi}}]{Matsuhira_2011}
\bibinfo{author}{\bibfnamefont{K.}~\bibnamefont{Matsuhira}},
  \bibinfo{author}{\bibfnamefont{M.}~\bibnamefont{Wakeshima}},
  \bibinfo{author}{\bibfnamefont{Y.}~\bibnamefont{Hinatsu}}, \bibnamefont{and}
  \bibinfo{author}{\bibfnamefont{S.}~\bibnamefont{Takagi}},
  \bibinfo{journal}{J. Phys. Soc. Jpn.}  (\bibinfo{year}{to be published}).

\bibitem[{\citenamefont{Koo et~al.}(2008)\citenamefont{Koo, Whangbo, and
  Kennedy}}]{Koo_1998}
\bibinfo{author}{\bibfnamefont{H.-J.} \bibnamefont{Koo}},
  \bibinfo{author}{\bibfnamefont{M.-H.} \bibnamefont{Whangbo}},
  \bibnamefont{and} \bibinfo{author}{\bibfnamefont{B.~J.}
  \bibnamefont{Kennedy}}, \bibinfo{journal}{J. Solid State Chem.}
  \textbf{\bibinfo{volume}{136}}, \bibinfo{pages}{269} (\bibinfo{year}{2008}).

\bibitem[{\citenamefont{Kim et~al.}(2008)\citenamefont{Kim, Jin, Moon, Kim,
  Park, Leem, Yu, Noh, Kim, Oh et~al.}}]{Kim_2008}
\bibinfo{author}{\bibfnamefont{B.~J.} \bibnamefont{Kim}},
  \bibinfo{author}{\bibfnamefont{H.}~\bibnamefont{Jin}},
  \bibinfo{author}{\bibfnamefont{S.~J.} \bibnamefont{Moon}},
  \bibinfo{author}{\bibfnamefont{J.-Y.} \bibnamefont{Kim}},
  \bibinfo{author}{\bibfnamefont{B.-G.} \bibnamefont{Park}},
  \bibinfo{author}{\bibfnamefont{C.~S.} \bibnamefont{Leem}},
  \bibinfo{author}{\bibfnamefont{J.}~\bibnamefont{Yu}},
  \bibinfo{author}{\bibfnamefont{T.~W.} \bibnamefont{Noh}},
  \bibinfo{author}{\bibfnamefont{C.}~\bibnamefont{Kim}},
  \bibinfo{author}{\bibfnamefont{S.-J.} \bibnamefont{Oh}},
  \bibnamefont{et~al.}, \bibinfo{journal}{Phys. Rev. Lett.}
  \textbf{\bibinfo{volume}{101}}, \bibinfo{pages}{076402}
  (\bibinfo{year}{2008}).

\bibitem[{\citenamefont{Hasegawa et~al.}(2010)\citenamefont{Hasegawa, Ogita,
  Matsuhira, Takagi, Wakeshima, Hinatsu, and Udagawa}}]{Hasegawa_2010}
\bibinfo{author}{\bibfnamefont{T.}~\bibnamefont{Hasegawa}},
  \bibinfo{author}{\bibfnamefont{N.}~\bibnamefont{Ogita}},
  \bibinfo{author}{\bibfnamefont{K.}~\bibnamefont{Matsuhira}},
  \bibinfo{author}{\bibfnamefont{S.}~\bibnamefont{Takagi}},
  \bibinfo{author}{\bibfnamefont{M.}~\bibnamefont{Wakeshima}},
  \bibinfo{author}{\bibfnamefont{Y.}~\bibnamefont{Hinatsu}}, \bibnamefont{and}
  \bibinfo{author}{\bibfnamefont{M.}~\bibnamefont{Udagawa}},
  \bibinfo{journal}{J. Phys.: Conf. Ser.} \textbf{\bibinfo{volume}{200}},
  \bibinfo{pages}{012054} (\bibinfo{year}{2010}).

\bibitem[{\citenamefont{Watahiki et~al.}(to be
  published)\citenamefont{Watahiki, Tomiyasu, Matsuhira, Iwasa, Yokoyama,
  Takagi, Wakeshima, and Hinatsu}}]{Watahiki_2011}
\bibinfo{author}{\bibfnamefont{M.}~\bibnamefont{Watahiki}},
  \bibinfo{author}{\bibfnamefont{K.}~\bibnamefont{Tomiyasu}},
  \bibinfo{author}{\bibfnamefont{K.}~\bibnamefont{Matsuhira}},
  \bibinfo{author}{\bibfnamefont{K.}~\bibnamefont{Iwasa}},
  \bibinfo{author}{\bibfnamefont{M.}~\bibnamefont{Yokoyama}},
  \bibinfo{author}{\bibfnamefont{S.}~\bibnamefont{Takagi}},
  \bibinfo{author}{\bibfnamefont{M.}~\bibnamefont{Wakeshima}},
  \bibnamefont{and} \bibinfo{author}{\bibfnamefont{Y.}~\bibnamefont{Hinatsu}},
  \bibinfo{journal}{J. Phys.: Conf. Ser.}  (\bibinfo{year}{to be published}).

\bibitem[{\citenamefont{Ohoyama et~al.}(1998)\citenamefont{Ohoyama, Kanouchi,
  Nemoto, Ohashi, Kajitani, and Yamaguchi}}]{HERMES_1998}
\bibinfo{author}{\bibfnamefont{K.}~\bibnamefont{Ohoyama}},
  \bibinfo{author}{\bibfnamefont{T.}~\bibnamefont{Kanouchi}},
  \bibinfo{author}{\bibfnamefont{K.}~\bibnamefont{Nemoto}},
  \bibinfo{author}{\bibfnamefont{M.}~\bibnamefont{Ohashi}},
  \bibinfo{author}{\bibfnamefont{T.}~\bibnamefont{Kajitani}}, \bibnamefont{and}
  \bibinfo{author}{\bibfnamefont{Y.}~\bibnamefont{Yamaguchi}},
  \bibinfo{journal}{Jpn. J. Appl. Phys.} \textbf{\bibinfo{volume}{37}},
  \bibinfo{pages}{3319} (\bibinfo{year}{1998}).

\bibitem[{\citenamefont{Kadowaki et~al.}(2002)\citenamefont{Kadowaki, Ishii,
  Matsuhira, and Hinatsu}}]{Kadowaki_2002}
\bibinfo{author}{\bibfnamefont{H.}~\bibnamefont{Kadowaki}},
  \bibinfo{author}{\bibfnamefont{Y.}~\bibnamefont{Ishii}},
  \bibinfo{author}{\bibfnamefont{K.}~\bibnamefont{Matsuhira}},
  \bibnamefont{and} \bibinfo{author}{\bibfnamefont{Y.}~\bibnamefont{Hinatsu}},
  \bibinfo{journal}{Phys. Rev. B} \textbf{\bibinfo{volume}{65}},
  \bibinfo{pages}{144421} (\bibinfo{year}{2002}).

\bibitem[{\citenamefont{Taira et~al.}(2001)\citenamefont{Taira, Wakeshima, and
  Hinatsu}}]{Taira_2001}
\bibinfo{author}{\bibfnamefont{N.}~\bibnamefont{Taira}},
  \bibinfo{author}{\bibfnamefont{M.}~\bibnamefont{Wakeshima}},
  \bibnamefont{and} \bibinfo{author}{\bibfnamefont{Y.}~\bibnamefont{Hinatsu}},
  \bibinfo{journal}{J. Phys.} \textbf{\bibinfo{volume}{13}},
  \bibinfo{pages}{5527} (\bibinfo{year}{2001}).

\bibitem[{com({\natexlab{a}})}]{comm_1}
\bibinfo{note}{The space group of undeformed pyrochlore structure is
  represented by $Fd\bar{3}m$ (No. 227), in which the atomic positions are
  described by (1/2, 1/2, 1/2) for Nd (16$d$), (0, 0, 0) for Ir (16$c$), ($x$,
  1/8, 1/8) for O (48$f$), and (3/8, 3/8, 3/8) for O
  (8$b$).~\cite{Gardner_2010} We also confirmed the same value $x=0.33(0)$ by
  analyzing the experimental integrated intensities of nuclear reflections at
  40 K for Nd$_2$Ir$_2$O$_7$.}

\bibitem[{\citenamefont{Yasui et~al.}(2001)\citenamefont{Yasui, Kondo, Kanada,
  Ito, Harashina, Sato, and Kakurai}}]{Yasui_2001}
\bibinfo{author}{\bibfnamefont{Y.}~\bibnamefont{Yasui}},
  \bibinfo{author}{\bibfnamefont{Y.}~\bibnamefont{Kondo}},
  \bibinfo{author}{\bibfnamefont{M.}~\bibnamefont{Kanada}},
  \bibinfo{author}{\bibfnamefont{M.}~\bibnamefont{Ito}},
  \bibinfo{author}{\bibfnamefont{H.}~\bibnamefont{Harashina}},
  \bibinfo{author}{\bibfnamefont{M.}~\bibnamefont{Sato}}, \bibnamefont{and}
  \bibinfo{author}{\bibfnamefont{K.}~\bibnamefont{Kakurai}},
  \bibinfo{journal}{J. Phys.} \textbf{\bibinfo{volume}{70}},
  \bibinfo{pages}{284} (\bibinfo{year}{2001}).

\bibitem[{\citenamefont{Freeman and Desclaux}(1979)}]{Freeman_1979}
\bibinfo{author}{\bibfnamefont{A.~J.} \bibnamefont{Freeman}} \bibnamefont{and}
  \bibinfo{author}{\bibfnamefont{J.~P.} \bibnamefont{Desclaux}},
  \bibinfo{journal}{J. Magn. Magn. Mater.} \textbf{\bibinfo{volume}{12}},
  \bibinfo{pages}{11} (\bibinfo{year}{1979}).

\bibitem[{\citenamefont{Marshall and Lovesey}(Oxford University Press, New
  York, 1971)}]{Marshall_1971}
\bibinfo{author}{\bibfnamefont{W.}~\bibnamefont{Marshall}} \bibnamefont{and}
  \bibinfo{author}{\bibfnamefont{S.~W.} \bibnamefont{Lovesey}},
  \bibinfo{journal}{{\it Theory of Thermal Neutron Scattering}}
  (\bibinfo{year}{Oxford University Press, New York, 1971}).

\bibitem[{\citenamefont{Matsuhira et~al.}(2002)\citenamefont{Matsuhira,
  Hinatsu, Tenya, Amitsuka, and Sakakibara}}]{Matsuhira_2002}
\bibinfo{author}{\bibfnamefont{K.}~\bibnamefont{Matsuhira}},
  \bibinfo{author}{\bibfnamefont{Y.}~\bibnamefont{Hinatsu}},
  \bibinfo{author}{\bibfnamefont{K.}~\bibnamefont{Tenya}},
  \bibinfo{author}{\bibfnamefont{H.}~\bibnamefont{Amitsuka}}, \bibnamefont{and}
  \bibinfo{author}{\bibfnamefont{T.}~\bibnamefont{Sakakibara}},
  \bibinfo{journal}{J. Phys. Soc. Jpn.} \textbf{\bibinfo{volume}{71}},
  \bibinfo{pages}{1576} (\bibinfo{year}{2002}).

\bibitem[{\citenamefont{Kobayashi et~al.}(2011)\citenamefont{Kobayashi, Nagao,
  and Ito}}]{Kobayashi_2011}
\bibinfo{author}{\bibfnamefont{K.}~\bibnamefont{Kobayashi}},
  \bibinfo{author}{\bibfnamefont{T.}~\bibnamefont{Nagao}}, \bibnamefont{and}
  \bibinfo{author}{\bibfnamefont{M.}~\bibnamefont{Ito}}, \bibinfo{journal}{Acta
  Cryst.} \textbf{\bibinfo{volume}{A67}}, \bibinfo{pages}{473}
  (\bibinfo{year}{2011}).

\bibitem[{\citenamefont{Tomiyasu et~al.}(2008)\citenamefont{Tomiyasu, Suzuki,
  Toki, Itoh, Matsuura, Aso, and Yamada}}]{Tomiyasu_2008}
\bibinfo{author}{\bibfnamefont{K.}~\bibnamefont{Tomiyasu}},
  \bibinfo{author}{\bibfnamefont{H.}~\bibnamefont{Suzuki}},
  \bibinfo{author}{\bibfnamefont{M.}~\bibnamefont{Toki}},
  \bibinfo{author}{\bibfnamefont{S.}~\bibnamefont{Itoh}},
  \bibinfo{author}{\bibfnamefont{M.}~\bibnamefont{Matsuura}},
  \bibinfo{author}{\bibfnamefont{N.}~\bibnamefont{Aso}}, \bibnamefont{and}
  \bibinfo{author}{\bibfnamefont{K.}~\bibnamefont{Yamada}},
  \bibinfo{journal}{Phys. Rev. Lett.} \textbf{\bibinfo{volume}{101}},
  \bibinfo{pages}{177401} (\bibinfo{year}{2008}).

\bibitem[{\citenamefont{Maegawa et~al.}(2010)\citenamefont{Maegawa, Oyamada,
  and Sato}}]{Maegawa_2010}
\bibinfo{author}{\bibfnamefont{S.}~\bibnamefont{Maegawa}},
  \bibinfo{author}{\bibfnamefont{A.}~\bibnamefont{Oyamada}}, \bibnamefont{and}
  \bibinfo{author}{\bibfnamefont{S.}~\bibnamefont{Sato}}, \bibinfo{journal}{J.
  Phys. Soc. Jpn.} \textbf{\bibinfo{volume}{79}}, \bibinfo{pages}{011002}
  (\bibinfo{year}{2010}).

\bibitem[{com({\natexlab{b}})}]{comm_2}
\bibinfo{note}{It should be reminded that {\it ferromagnetic} interactions
  generate the highly frustrated spin-ice states (2-in 2-out) of anisotropic
  moments in a pyrochlore lattice.~\cite{Bramwell_2001}}.

\bibitem[{\citenamefont{Kim et~al.}(2009)\citenamefont{Kim, Ohsumi, Komesu,
  Sakai, Morita, Takagi, and Arima}}]{Kim_2009}
\bibinfo{author}{\bibfnamefont{B.~J.} \bibnamefont{Kim}},
  \bibinfo{author}{\bibfnamefont{H.}~\bibnamefont{Ohsumi}},
  \bibinfo{author}{\bibfnamefont{T.}~\bibnamefont{Komesu}},
  \bibinfo{author}{\bibfnamefont{S.}~\bibnamefont{Sakai}},
  \bibinfo{author}{\bibfnamefont{T.}~\bibnamefont{Morita}},
  \bibinfo{author}{\bibfnamefont{H.}~\bibnamefont{Takagi}}, \bibnamefont{and}
  \bibinfo{author}{\bibfnamefont{T.}~\bibnamefont{Arima}},
  \bibinfo{journal}{Science} \textbf{\bibinfo{volume}{323}},
  \bibinfo{pages}{1329} (\bibinfo{year}{2009}).

\bibitem[{\citenamefont{Gardner et~al.}(2010)\citenamefont{Gardner, Gingras,
  and Greedan}}]{Gardner_2010}
\bibinfo{author}{\bibfnamefont{J.}~\bibnamefont{Gardner}},
  \bibinfo{author}{\bibfnamefont{M.~J.~P.} \bibnamefont{Gingras}},
  \bibnamefont{and} \bibinfo{author}{\bibfnamefont{J.~E.}
  \bibnamefont{Greedan}}, \bibinfo{journal}{Rev. Mod. Phys.}
  \textbf{\bibinfo{volume}{82}}, \bibinfo{pages}{53} (\bibinfo{year}{2010}).

\end{thebibliography}

\end{document}